\documentclass[showpacs,preprintnumbers,amsmath,amssymb,preprint]{revtex4}
\usepackage{graphics}

\begin{document}

\title{Break-up stage restoration in multifragmentation reactions}

\small
\author{Ad. R. Raduta$^{1,2}$, E. Bonnet$^{1}$, B. Borderie$^{1}$,
N. Le Neindre$^{1}$, S. Piantelli$^{3}$ and M. F. Rivet$^{1}$}
\address{
$^{1}$Institut de Physique Nucleaire, IN2P3-CNRS, F-91406 Orsay cedex, France\\
$^{2}$NIPNE, Bucharest-Magurele, POB-MG 6, Romania\\
$^{3}$Dip. di Fisica e Sezione INFN, Universit\`a di Firenze, I-50019
Sesto Fiorentino (Fi), Italy
}

\begin{abstract}
In the case of Xe+Sn at 32 MeV/nucleon multifragmentation reaction
break-up fragments are built-up from the experimentally detected ones
using evaluations of light particle evaporation multiplicities
which thus settle fragment internal excitation.
Freeze-out characteristics are extracted from
experimental kinetic energy spectra
under the assumption of full decoupling between fragment formation and
energy dissipated in different degrees of freedom. 
Thermal kinetic energy is determined uniquely while
for freeze-out volume - collective energy a multiple solution is obtained.
Coherence between the solutions of the break-up restoration algorithm and
the predictions of a 
multifragmentation model with identical definition of
primary fragments is regarded as a way to select the true value.
The broad kinetic energy spectrum of $^3$He is consistent with break-up
genesis of this isotope. 
\end{abstract}

\maketitle

\section{Introduction}

The possibility to investigate the phase diagram of highly excited nuclear
systems made multifragmentation one of the most lively debated subjects in
nowadays physics.
However, the approach of the thermodynamically relevant break-up stage of the
process is difficult from both conceptual and practical perspectives.
From the theoretical point of view, the break-up definition of an open system
requires reconsideration of a fundamental thermodynamical quantity,
the volume \cite{gulminelli}.
From the experimental point of view, all measured observables
are perturbed by non-equilibrium phenomena and
subsequent disintegration of primary excited fragments.

Adopting the freeze-out equilibrium hypothesis,
the purpose of the present work is to investigate to what extent the break-up stage
may be restored from limited but relevant experimental data.
The high quality of the INDRA detector and the wealth of already existing analyses
\cite{frankland1,frankland2,hudan}
that indicate the formation of an equilibrated source
make the Xe+Sn at 32 MeV/nucleon multifragmentation reaction the most suitable
candidate for such a study.
The present paper follows the line opened by Piantelli {\it et al.} in Ref. \cite{piantelli}
whose goal was that of estimating 
freeze-out properties such as the volume in
a fully consistent and model-independent way.
Quite remarkably, if constrained by all available experimental information,
the requirement of optimal fitting of both
measured light charged particle and fragment kinetic spectra resulted in rather
high average excitation energy of primary fragments (3.9 MeV per nucleon).
While both stick to the break-up concept, 
the present work distinguishes from Ref. \cite{piantelli} by considering that
break-up partitions can be traced-back using
an evaluation of evaporated particle multiplicities \cite{hudan} and
treating all energetic degrees of freedom as fully decoupled.

After a brief review of experimental information, Section II describes the algorithm
designed to determine break-up fragment partition and average excitation energy.
Attempts to identify the freeze-out volume ($V$) and kinetic properties 
(thermal ($K_{th}$) and collective ($E_{coll}$) energies and flow profile $a_{coll}$)
of the break-up stage are presented in Section III.
Section IV discusses the uniqueness of the solution and coherence with results
of a multifragmentation model with identical fragment definition
is thought as a way to select the "real" value and
to check the consistency of the results.
In Section V we review the steps-forward accomplished by different
studies meant to characterize the break-up stage of the Xe+Sn at 32 MeV/nucleon
multifragmentation reaction,
their limitations and the relevance of present results.

\section{Determining break-up fragment partition and excitation energy}

Experimental data files corresponding to Xe+Sn at 32 MeV/nucleon multifragmentation
reaction \cite{frankland1,frankland2} are used to reconstruct the thermalized source
formed in central collisions. 
As a first step, quasi-complete events where total detected charge exceeds 80\%
of the summed up charge of the target and projectile are selected and
the emission isotropy for fragments
is ensured by large values of the fragment flow angle,
$\theta_{flow} \geq 60^o$.
By correcting for detection efficiency, the cross section of
these selected isotropic central collisions is estimated at 115 $\pm$ 20 mb
\cite{hudan}. 
Mass of the isotopically un-resolved products with $Z \geq 5$ is calculated
assuming an EAL dependence \cite{charity}. 
Then, light charged particles ($1 \leq Z \leq 4 $) emitted at
forward ($\theta_{CM} \leq 60^0$) and backward ($\theta_{CM} \geq 120^0$)
angles in the center of mass of the reaction are rejected as partly created
at pre-equilibrium.
Therefore, light charged particles emitted outside this rejection zone
are doubled under the assumption of isotropic emission.
Neutrons (which are undetected) are added to each event to conserve
the $N/Z$ of the total system.
At this stage the source characterized by $<A>$=198 and $<Z>$=83 is considered
equilibrated and break-up fragments are built-up from the detected ones
attributing randomly neutrons and light charge particles to
heavier fragments ($Z \geq 5$) according to
pre-defined evaporated particle multiplicities
and weighted by final sizes of fragments.
For the percentages of evaporated 
deuterons, tritons, $^4$He and $^6$He we used the following values:
$p_{\rm d}$=0.36, $p_{\rm t}$=0.46,
$p_{^4{\rm He}}$=0.60 and $p_{^6{\rm He}}$=0.30
obtained using particle-fragment correlation measurements \cite{hudan}
and correcting for pre-equilibrium using the method proposed in
Ref. \cite{frankland2}.
The percentage of evaporated protons $p_{\rm p}$ is set to 0.30
assuming that the contribution of fragments with $Z>35$,
not considered in Ref. \cite{hudan},
modifies exclusively the proton yield.
The percentages of evaporated
neutrons and nuclei with $Z=3,4$ are free parameters which have to be determined
simultaneously with the average excitation energy of the primary products such as
to recover, after sequential particle evaporation,
the experimental charge distribution.
The apart case of $^3$He whose average kinetic energy exceeds by 15 MeV that of
tritons and by 13 MeV that of $^4$He and the urge of not introducing extra
hypothesis in addition to the break-up scenario, make us tentatively assume
that $^3$He originates prevalently
from the break-up stage ($p_{^3{\rm He}}$=0) \cite{nicolas}.
Though not relevant for reconstruction of break-up fragment partition
due to the relatively low multiplicity of $^3$He ($Y(^3He)$=0.9),
{\em if true},
this assumption {\em may} provide an additional test for the accuracy of
break-up kinetic properties determination,
as we shall see later on.

We recall that the paradoxical behavior of $^3$He is already notorious as
evidenced in a multitude of nuclear reactions covering a broad energy domain
({\it eg.} 7.5 GeV/c p+$^{12}$C,$^{112,124}$Sn \cite{bogatin},
Ne+U at 250 and 400 MeV/nucleon \cite{gutbrod},
202 MeV/c $\bar p$+$^{12}$C, $^{40}$Ca,$^{63}$Cu, $^{92,98}$Mo,$^{238}$U \cite{markiel},
central collisions of Au+Au at 100, 150 and 250 MeV/nucleon \cite{poggi}
and Xe+Sn at 50 MeV/nucleon \cite{bougault})
and impossible to understand within a thermal scenario 
(according to which $<K>_{th}(^3{\rm He} ) \approx <K>_{th}(^4{\rm He} )$)
eventually coupled with a collective expansion 
(providing $<K>_{coll}(^3{\rm He} ) < <K>_{coll}(^4{\rm He} )$).
The high kinetic energy of $^3$He was so far explained as a signature of its
early synthesis from coalescing non-equilibrium nucleons \cite{neubert}
or a time expanding source \cite{bougault}.
As the coalescence as a mechanism for fragment formation is expected to work
for energies higher than the one which characterizes our reaction and the
fragment emission from an expanding source \cite{ees}
is incompatible with the adopted break-up scenario,
the assumption of the break-up genesis of $^3$He,
in contradistinction to other charged particles and fragments which may originate
from both break-up and evaporation stages, seems to be the most appropriate.

\begin{figure}
\resizebox{0.6\textwidth}{!}{%
  \includegraphics{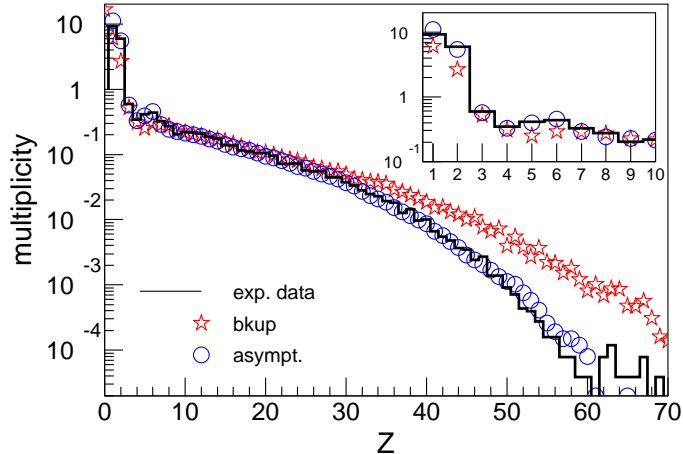}
  }  
\caption{
Reconstructed break-up and asymptotic charge distributions corresponding to the
(198,83) equilibrated source formed in Xe+Sn at 32 MeV/nucleon multifragmentation
reaction in comparison with experimental data.}
\label{fig:yz_exp+bkup}
\end{figure}

So, assuming for the primary fragments a roughly constant excitation energy
per mass unit ($\epsilon$) \cite{hudan,tlim},
we allow these fragments to de-excite
following a standard Weisskopf scheme where realistic evaporation Coulomb barriers
\cite{rivet,vaz}
and complete experimental information on excited states of nuclei with $A<14$
have been implemented.
The value of $\epsilon$ which reproduces
the experimental charge multiplicity within 10\% is 2.5 MeV/nucleon, 
in good agreement with the value proposed by Hudan {\it et al.},
2.2 MeV/nucleon \cite{hudan}.
The "reconstructed" break-up and asymptotic charge distributions are presented in
Fig. \ref{fig:yz_exp+bkup} in comparison with the experimental data.
For the percentage of evaporated neutrons and nuclei with $Z=3,4$ we
obtained 0.40, 0.14 and 0.10, respectively.

To complete the static picture of the break-up stage we mention that
the average values of total multiplicity ($N_{tot}$),
charged product multiplicity ($N_{c}$)
and fragment multiplicity ($N_{Z \geq 5}$)
are 30.5 (asymptotic 53.8),
13.8 (asymptotic 23.2; experimental 20.3)
and 4.2 (asymptotic 4.0; experimental 4.2) respectively.
The average total excitation energy ($E_{ex}=\sum_i \epsilon_i$) is 
396.9 MeV and Q-value equals 126.7 MeV.
Both total multiplicity and Q-value are subject to variation if final fragments
isotopic composition differs from the presently assumed one.

\section{Attempts to determine the freeze-out volume and
kinetic properties of the break-up configuration}

The restoration of the fragment configuration event by event suggests
the possibility to determine
the average values of the other observables which characterize the break-up
(the freeze-out volume 
and the thermal and collective energies)
by comparing the simulated asymptotic kinetic energy distributions
$<K>$ vs. $Z$ and $dN/dK$ vs. $K$ with the experimental ones.
The idea of extracting information on the freeze-out stage and in particular on
the freeze-out volume using the energy gain during Coulomb propagation
was advanced for the first time in Refs. \cite{li,sangster,bb}
under the simplifying assumptions of low excitation of primary fragments
and negligible radial collective flow.
To reach our goal, we assume full decoupling between fragment formation at freeze-out
and amounts of energy dissipated in different degrees of freedom
and define the freeze-out volume as the spherical container
in which the primary fragments are localized.
Interesting enough, present consideration of volume agrees equally with
the two extreme scenarios in the literature: volume as an external constraint as
routinely employed by microcanonical models \cite{statmod} or
simply the spatial extension of the expanding system when fragments
cease to interact by nuclear force
or final multiplicity is attained as in dynamical models \cite{dynmod}.

\begin{figure}
\resizebox{0.6\textwidth}{!}{%
  \includegraphics{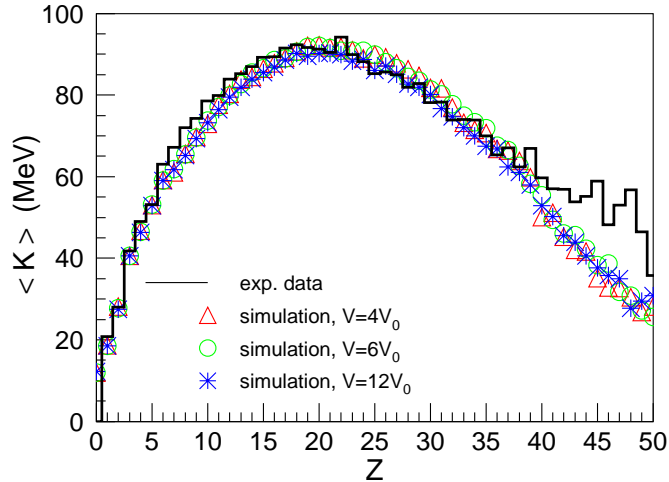}}  
\caption{
Asymptotic simulated average kinetic energy distributions
as a function of fragment charge corresponding to
different values of freeze-out volume and collective energy
in comparison with experimental data of the
Xe+Sn at 32 MeV/nucleon multifragmentation reaction.
Complete particle evaporation was performed after full propagation, at 500 fm/c.}
\label{fig:kz_exp+as}
\end{figure}

\begin{figure}
\resizebox{0.9\textwidth}{!}{%
  \includegraphics{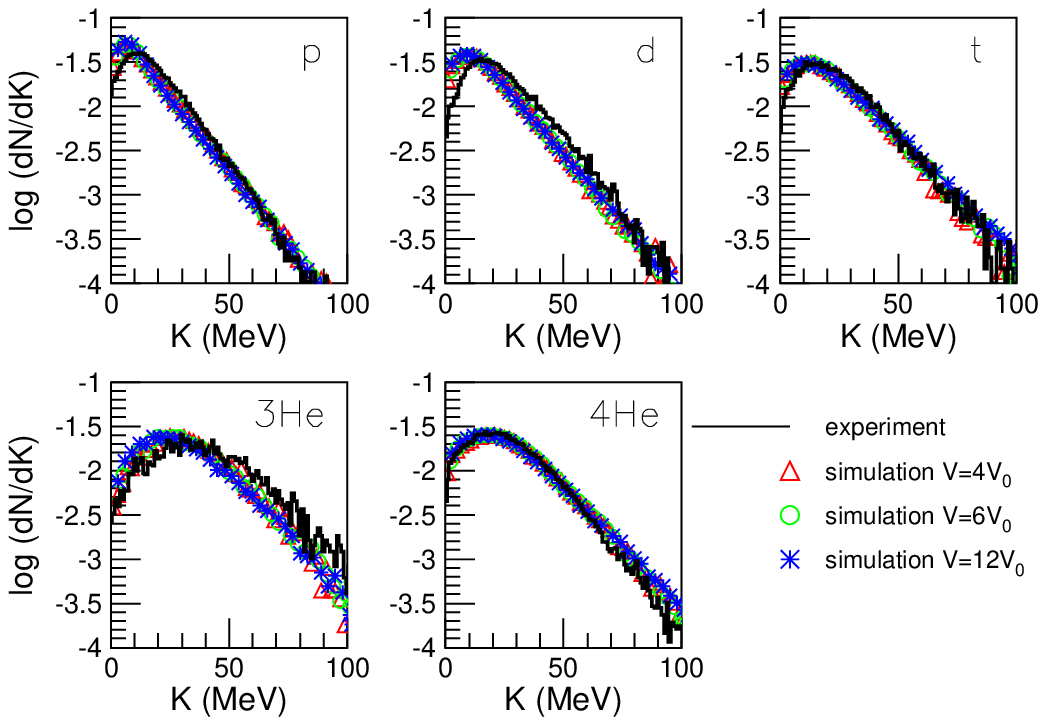}}  
\caption{
Asymptotic simulated kinetic energy spectra of light charged particles
corresponding to
different values of freeze-out volume and collective energy
in comparison with experimental data of the
Xe+Sn at 32 MeV/nucleon multifragmentation reaction.
Complete particle evaporation was performed after full propagation, at 500 fm/c.}
\label{fig:dndk_exp+as}
\end{figure}

For simplicity, fragments are treated as normal nuclear density
non-overlapping spheres,
their size being thus the experimentally measured one if $Z \leq 6$  \cite{radii}
or calculated according to $R=1.2 A^{1/3}$ for heavier nuclei and
the freeze-out volume is assimilated with the smallest sphere which includes entirely
the fragments.
Once localized, fragments share the available thermal kinetic energy, 
a free parameter in our simulation,
according to a Maxwell distribution
under total momentum and angular momentum conservation laws
and then propagate in the free space under the action of collective
motion and Coulomb repulsion. 
Several values of the freeze-out volume have been considered, ranging from 4$V_0$
- the lowest value allowed by the fragment non-overlapping condition
up to 12$V_0$ - slightly exceeding the maximum evaluations
in the literature \cite{dynmod,mmm},
where $V_0$ stands for the volume at normal nuclear density.
Taking into account that recent experimental works \cite{lefevre} indicate
for the flow profile values larger than 1, we consider this quantity
as a free parameter.
In each case average total thermal energy $K_{th}$, magnitude $E_{flow}$ and
profile $a_{flow}$ of collective flow
have been tuned to provide the best agreement between the simulated asymptotic
$<K>$ vs. $Z$ and $dN/dK$ vs. $K$ distributions and the corresponding experimental data.
The results indicate that Coulomb and flow energy, both dependent on fragment size and
position, may compensate each other so that equally good results are produced by
any suitably tuned (freeze-out volume, collective energy) set.
This result is visible in Figs. \ref{fig:kz_exp+as} and \ref{fig:dndk_exp+as}
where the asymptotic $<K>$ vs. $Z$ and $dN/dK$ vs. $K$ distributions
of light charged fragments are plotted for three situations:
(i) $V=4V_0$, $E_{flow}$=0.68 MeV/nucleon, $a_{flow}$=1.3;
(ii) $V=6V_0$, $E_{flow}$=0.9 MeV/nucleon, $a_{flow}$=1.2 and
(iii) $V=12V_0$, $E_{flow}$=1.24 MeV/nucleon, $a_{flow}$=1.2.
As one may notice from Fig. \ref{fig:kz_exp+as}, the agreement between the 
asymptotic stage simulated $<K>$ vs. $Z$ distribution and the
experimental data is good up to $Z$=40, while serious underestimation is
produced for heavier fragments. The reason may be related to the arbitrary
way in which break-up fragments larger than $Z$=35 have been built-up
from the experimental partition in lack of any estimation of the
evaporated particle multiplicities.
In what regards the $dN/dK$ vs. $K$ distributions, in general
the simulated spectra of light charged particles describe well
the experimental ones (see Fig. \ref{fig:dndk_exp+as}),
while systematic underestimation
(of about 70 \%(for $V=4V_0$) - 80 \% (for $V=12V_0$)) 
of the observed widths 
is realized for fragments (not shown).
 
The only kinetic quantity whose univocal determination was possible following
the above procedure is the total thermal energy as it influences strongly
the slope of the decreasing tail of the kinetic energy spectra of
light charged particles which is
practically insensitive to Coulomb repulsion and collective motion
in the considered domains.
The obtained value is $K_{th}$=396 MeV and corresponds to
a kinetic temperature, $T=K_{th}/(3/2 N_{tot})$, of 8.65 MeV.
We stress here that the occurence of nearly identical values
for $K_{th}$ and $E_{ex}$ is fortuitous and should not be assigned a physical meaning.
In what regards the $dN/dK$ vs. $K$ distributions of fragments,
the results of our simulation indicate significant and complex dependence
on all input quantities ($K_{th}$, $E_{flow}$, $V$), a quantification
of their effects being impossible.


It is interesting to notice here that
the simultaneous good description
of the kinetic energy spectra of $^3$He and
of those of p, d, t and $^4$He
supports the hypotheses of (prevalent) break-up formation
of this nucleus and, at the same time,
may be regarded as a validity test
of the procedure employed to determine $K_{th}$.

\begin{figure}
\resizebox{0.9\textwidth}{!}{%
  \includegraphics{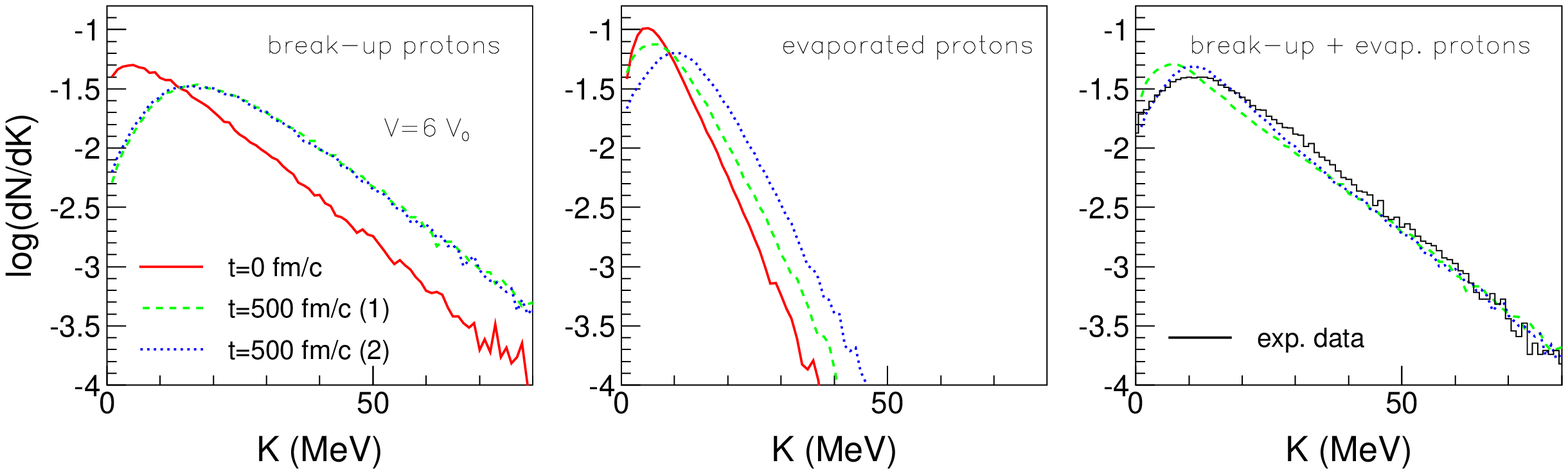}}  
\caption{Simulated kinetic energy spectra 
corresponding to $V=6 V_0$, $E_{coll}$=0.9 MeV/nucleon and $a_{coll}$=1.2
for the break-up (left),
evaporated (middle) and break-up + evaporated protons (right) when
sequential evaporation occurs at 500 fm/c (1) and,
respectively, 0 fm/c (2).
In the third panel the experimental distribution is plotted with histogram.
In the first panel the final distributions corresponding to
the two evaporation instances (dotted and dashed lines) nearly coincide
pointing out the modest increase of the Coulomb energy
caused by an early evaporation which practically takes place
in the freeze-out volume.}
\label{fig:tevap}
\end{figure}

A criticism to our procedure may concern the stability of the solution with respect to
the moment when excited fragments suffer sequential evaporation.
To clarify this issue kinetic energy spectra have been compared in two extreme cases:
(1) full particle evaporation occurs after complete propagation, at 500 fm/c 
and (2) full particle evaporation takes place just after break-up, at 0 fm/c.
In the second case, the final fragments are propagated
in the correspondingly altered Coulomb field during 500 fm/c.
In particular, the relatively advanced fragmentation of the system at break-up
and the rather poor particle production during sequential evaporation determine
an increase of the total Coulomb energy by only 1 \% in the case of early evaporation
with respect to the break-up stage.
The most important effect was evidenced for protons and the
$dN/dK$ vs. $K$ distributions are plotted in Fig. \ref{fig:tevap}
for the particular case of
$V=6 V_0$, $E_{coll}$=0.9 MeV/nucleon and $a_{coll}$=1.2.
As one may see in the third panel,
the two distributions differ only in the low energy domain,
the decreasing tail used to determine $K_{th}$ being practically unmodified.
The case corresponding to early de-excitation gives a better reproduction
of the experimental data and an average kinetic energy of $Z=1$ particles
exceeding by 1.6 MeV the result predicted by the
opposite case.
The influence of the evaporation instant on the kinetic energy spectra
may be understood taking into account the behavior of the
break-up and evaporated proton spectra.
Thus, the already discussed poor sensitivity of the break-up protons kinetic energy
to moderate variations of the Coulomb field
leads to perfectly superimposable $dN/dK$ vs. $K$ distributions corresponding
to the two evaporation scenarios.
For the sake of completeness, the left panel of Fig. \ref{fig:tevap} depicts also
the $dN/dK$ vs. $K$  distribution of protons at break-up (t=0 fm/c).  
Apart the obvious increase of the kinetic energy,
the propagation in the Coulomb field
produces a partial suppression of the low energy range.
In what regards the evaporated protons, the middle panel of Fig. \ref{fig:tevap}
indicates a significant increase of the kinetic energy in the case of
early evaporation with respect to the (1) scenario.
However, the widths of these distributions determined mainly by the break-up fragments
excitation energy are small enough so that only the low energy range
of the total $dN/dK$ vs. $K$  distributions proves sensitive to the evaporation time
(right panel of Fig. \ref{fig:tevap}).
 
Once estimated the average energy dissipated
in different degrees of freedom, one may evaluate the
total available energy of the source.
Thus, the calorimetric equation,
\begin{equation}
E_{tot}=Q+E_{ex}+K_{th}+V_{Coulomb}+E_{flow},
\end{equation}
predicts for all considered situations a total 
energy of 7.4 MeV/nucleon,
the 
$E^*$=($Q+E_{ex}+K_{th}+V_{Coulomb}$) part ranging
from 6.15 MeV/nucleon at 12$V_0$ to 6.7 MeV/nucleon at 4$V_0$.


\section{Additional information on the break-up stage:
comparison with a multifragmentation model}

The limitations of the above procedure to
completely characterize the break-up stage
of a nucleus using all available experimental data should not be attributed
to the technical flaws of the adopted recipe
but rather reflect the simplified treatment of
the phenomenon and require further consideration. 
Thus, the break-up is defined as the ultimate stage of the cooling-expansion
process suffered by the hot and initially compressed matter formed in 
heavy ion collisions
when it ceases to exist as a mononuclear configuration and crashes into pieces.
With this picture in mind, it is natural to admit that all break-up observables
should depend on the initial energy transferred into the system.
Recent results obtained within a microscopic approach \cite{de}
are interesting in this sense as they
show in the pre-break-up stage a monotonic increase of both
nuclear volume and expansion energy with the excitation energy.
On the other hand, statistical models which compute the weights of
all possible fragment configurations
$C:\{\{A_i,Z_i,\epsilon_i,{\rm\bf r}_i,{\rm\bf p}_i\}, i=1,...,NF\}$
compatible with a certain state of the equilibrated source
(eg. for microcanonical models, ($A,Z,E,V$))
show a strong interplay between all energetic degrees of freedom, including $Q$.
In this perspective, one possibility to select the unique physical break-up state
from the multiple solutions obtained by the restoration algorithm
is to perform a consistency test with predictions of a multifragmentation model,
statistical or dynamical.
Fragments treatment as hard non-overlapping spheres characterized by a normal-density
zero-temperature binding energy which may be additionally excited
together with a freeze-out volume configured as a spherical container 
in which
break-up fragments are localized recall the break-up stage description of the
MMM version \cite{mmm} of the microcanonical multifragmentation models
and encourage comparison with the predictions of this model.
We recall that, nevertheless, within the microcanonical models, the freeze-out volume
has more dramatic consequences as it affects not only the kinetic energy spectra, 
but the fragment partitions themselves via geometric restrictions and, again, 
Coulomb energy which this time enters the statistical weight of a configuration 
via energy balance \cite{mmm}.
Finally, as only mean values will be considered along this paper, it is reasonable
to assume that the freeze-out volume is fixed, without refuting the fact that
in experiment it may in principle vary from one event to another.
A reasonable agreement may constitute an additional proof in favor of
the break-up scenario.

Therefore, we start from the same equilibrated source
($A_0$,$Z_0$)=(198,83) and consider for
the freeze-out volume and excitation energy the values resulted from
the break-up reconstruction algorithm:
($V=4V_0$, $E^*$=6.7 MeV/nucleon),
($V=6V_0$, $E^*$=6.5 MeV/nucleon) and
($V=12V_0$, $E^*$=6.15 MeV/nucleon).
Because for microcanonical models
the break-up stage characteristics were proved to depend dramatically
on primary fragment definition in terms of excitation energy \cite{tlim},
two situations will be addressed.
In both cases fragments are allowed to absorb excitation up to the binding level,
but in the first case the high energy decrease of the nuclear level density,
\begin{equation}
\rho(\epsilon)=\frac{\sqrt{\pi}}{12 a^{1/4}\epsilon^{5/4}}
~ \exp(2 \sqrt{a \epsilon}) ~ \exp(-\epsilon/\tau),
\label{eq:rho}
\end{equation}
with  $a=0.114 A+0.098 A^{2/3}$ MeV$^{-1}$ 
is modulated
by a constant limiting temperature factor ($\tau$=9 MeV) \cite{iljinov}, 
while a mass dependent one is used in the second case
($\tau(A)=4.3+18.3/\sqrt{A}-19.9/A$ MeV) \cite{natowitz,baldo}.
Although finite values of $\tau$
result in different values of internal and kinetic temperature, 
this does not represent an equilibrium violation \cite{npa2002}. 
The presence of a high energy cut-off on the level density is 
required by the fragments binding criterion and is a common feature
of models which use clusters and not particles as degrees of freedom.

\begin{figure}
\resizebox{0.5\textwidth}{!}{%
  \includegraphics{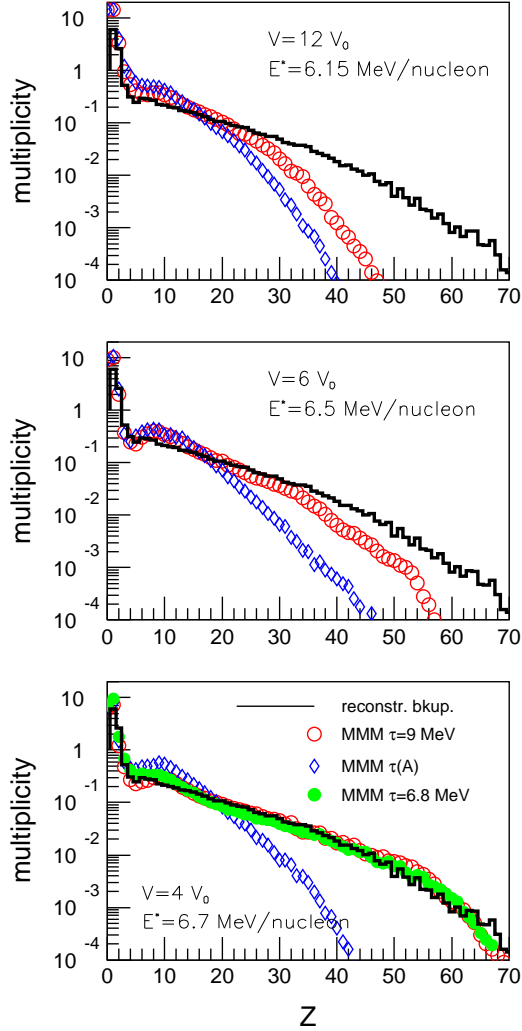}}  
\caption{Break-up stage fragment charge multiplicity distributions obtained by MMM 
for different states of the source ($A_0$,$Z_0$)=(198,83)
as indicated on the figure
in comparison with the experimental reconstructed distribution
corresponding to Xe+Sn at 32 MeV/nucleon multifragmentation reaction.
}
\label{fig:yz_bkup+mmm}
\end{figure}

Table I summarizes the values of different partial energies 
and product multiplicities at break-up
obtained in each considered case,
while Fig. \ref{fig:yz_bkup+mmm} depicts the
charge multiplicity distributions produced by MMM against
the reconstructed experimental break-up distribution.

\begin{table}
\begin{center}
\caption{Break-up stage values of average total thermal kinetic energy, excitation energy,
Coulomb energy, $Q$ value and excitation energy per mass unit
of primary fragments
expressed in MeV and total multiplicity, 
charged product multiplicity and fragment multiplicity
obtained by MMM for the multifragmentation of the (198, 83) source with
different freeze-out volumes and excitation energies and two definition of primary
fragments in terms of excitation energy, as indicated in the first column
in comparison with the reconstructed experimental break-up stage values.}
\begin{tabular}{ccccccccc}
\hline
\hline
($V/V_0$, $E^*$ (MeV),$\tau$ (MeV))&$K_{\rm th}$&$E_{\rm ex}$ &$V_{\rm Coulomb}$ & $Q$ & $\epsilon/A$ 
& $N_{tot}$& $N_{ch}$& $N_{Z \geq 5}$ \\
\hline
(12, 6.15, 9)         & 358.1 & 269.8 & 295.7 & 294.0 & 2   & 37.9 & 23.4 & 4.8 \\
(12, 6.15, $\tau(A)$) & 369.9 & 196.6 & 310.4 & 340.7 & 1.5 & 40.0 & 25.4 &5.4\\
(6, 6.5, 9)           & 321.7 & 425.8 & 358.8 & 180.6 & 2.7 & 26.6 & 17.5 & 5.1 \\
(6, 6.5, $\tau(A)$)   & 346.5 & 307.6 & 388.0 & 244.8 & 1.8 & 29.2 & 20.0 & 6.2 \\
(4, 6.7, 9)           & 282.4 & 565.6 & 382.2 & 96.3 & 3.2 & 19.5 & 13.4 & 4.8 \\ 
(4, 6.7, $\tau(A)$)   & 316.0 & 408.3 & 435.4 & 166.8 & 2.2 & 22.1 & 16.0 & 6.6 \\
(4, 6.7, 6.8)         & 371.3 & 401.1 & 394.7 & 159.5 & 2.5 & 23.6 & 16.5 & 4.7 \\
\hline
reconstr. exp. break-up & 396.0 & 396.9 & - & 126.7 & 2.5 &  30.5 & 13.8 & 4.2 \\
\hline
\hline
\end{tabular}
\end{center}
\label{table:en}
\end{table}

As one may notice,
contrary to what we have seen above, microcanonical models prove extremely sensitive
to the freeze-out volume - excitation energy ($E^*$) correlated variation.
By favoring asymmetric configurations with lower fragment multiplicity,
a decreasing freeze-out volume leads to a broadening of the charge distribution
and the drop of the Q-value. Quite interesting is the fact that
while the evolution of product multiplicities and partial energies is similar
for the two considered hypotheses of internal excitation,
the effect on $Y(Z)$ distributions differs significantly.
Thus, by increasing the freeze-out volume,
a constant limiting temperature produces a reduction of the charge distribution range
by 20 units (open circles),
while for a mass dependent $\tau$ the modification is ten times smaller 
(open diamonds) \cite{tlim}.
In what regards the variation of total Coulomb repulsion,
the values in Table I indicate that the effect of a more uniform population accomplished
for large volumes due to an increased fragmentation
is annihilated by the influence of the volume size such that
\textbf{an} increasing volume produces a monotonic decrease of $V_{Coulomb}$.
Finally, according to MMM more expanded configurations are characterized
by relatively colder fragments ($\epsilon$=1.5 - 2 MeV/nucleon at 12$V_0$ and
$\epsilon$=2.2 - 3.2 MeV/nucleon at 4$V_0$)
with correspondingly lower total internal excitation.
As in microcanonical models thermal kinetic energy is determined
event by event by the available energy,
the variation of $K_{th}$ results from the interplay between
the other energetic degrees of freedom and,
as indicated in table I, it manifests a monotonic decrease with the volume.

Turning back to the possibility to identify the break-up stage
performing a consistency test with the predictions of MMM,
one may say that even if for $V=4V_0$, $E^{*}$=6.7 MeV/nucleon and $\tau$=9 MeV
the calculated $Y(Z)$ distribution agrees well with the
reconstructed experimental one,
the result is unacceptable due to the
considerably different sharing of total energy:
$K_{th}$ and Q are underestimated by 30\% and, respectively, 24 \%
and $E_{ex}$ is overestimated by 40 \%.
An extra diminishing of the freeze-out volume
can not be considered a solution as it will lead to $Y(Z)$ distributions broader
than the experimental one and $K_{th}$, $E_{ex}$ and $Q$ will be still worse.
The amount of energy dissipated in $K_{th}$, $E_{ex}$ and $Q$ in better agreement
with the estimations of the break-up reconstruction algorithm obtained for
$V=4V_0$ and $E^{*}$=6.7 MeV/nucleon when $\tau(A)$ ranges from
5.9 (for $A$=100) to 8.5 (for $A$=5)
suggests as a possible solution of our problem
the reduction of the constant value of the limiting temperature.
Thus, diminishing $\tau$ to the arbitrary value 6.8 MeV,
we get an $Y(Z)$ distribution which follows closely the "experimental" break-up curve and
partial energies in acceptable agreement with the "measured" ones.
The only quantity for which MMM and the break-up reconstruction algorithm give
different results is the total multiplicity and the explanation
relies on the different isotopic composition of the primary fragments in the two cases.
Indeed, in MMM the break-up products are rather neutron rich and
relatively few neutrons are created at break-up. 
The good reproduction of kinetic distributions $<K>$ vs. $Z$ and $dN/dK$ vs. $K$
after sequential evaporation does not require further consideration as
it is guaranteed by the procedure adopted to build the break-up stage
from experimental data.

\section{Discussion and conclusions}

Even if promising as an approximate determination of an average freeze-out volume
consistent with some relevant experimental information,
present results depend upon fragment definition,
further studies on the shape, density and related internal excitation of nuclei at
break-up being mandatory for a definite answer.
For instance, the ambitious task of checking the consistency between
statistical and dynamical model predictions is expected to provide information
on fragment characteristics at break-up and answer the fundamental
question whether full equilibration is attained in heavy ion reactions.
A remarkable attempt in this sense, 
though restricted to fragment partitions and
average kinetic energy as a function of charge
and dealing with a slightly different source,
has been recently published in Ref. \cite{mmm+smf} reaching
the conclusion that a distinct equilibrated stage may be identified in the dynamical
evolution of the presently considered Xe+Sn at 32 MeV/nucleon reaction as
described by the Stochastic Mean Field approach \cite{smf}.
 
The difficulty to restore the break-up stage using exclusively asymptotic
experimental information and
the important role played by fragments internal excitation
are well illustrated by Refs. \cite{piantelli,tlim}. 
For instance, Ref. \cite{tlim} provides an example on how 
the agreement between experimental average properties of fragments and
asymptotic stage predictions of a statistical multifragmentation model
as criterion for identifying the equilibrated state of a nuclear source
may provide for the freeze-out volume values differing by a factor of two when
various parameterizations of the break-up fragments limiting temperature
are employed.
Even more eloquent is the case of Ref. \cite{piantelli} where
the assumed relation between the thermal kinetic temperature,
break-up fragment internal temperature and limiting temperature
($1/T_{frag}=3/2<K_{fo}>^{-1}+1/\tau$) \cite{randrup-tlim}
fixes the sharing between fragment excitation and thermal energy.
This fact, coupled to some differences in the building of the 
freeze-out stage such as the distance between fragments' surfaces, 
which in the present study may approach zero,
determines the 
variance between break-up fragments characteristics obtained in 
Ref. \cite{piantelli} with respect to the present work.

In conclusion, 
for Xe+Sn at 32 MeV/nucleon multifragmentation reaction
fragment partitions at break-up were built event by event
from the experimental fragment partitions using experimentally evaluated
average multiplicities of evaporated light charged particles
which also allow implicit estimation of the limited average excitation energy
of primary fragments.
Kinetic characteristics of the break-up stage were inferred from the comparison
of simulated asymptotic $<K>$ vs. $Z$ and $dN/dK$ vs. $K$ distributions
with the experimental ones. 
$K_{th}$ was determined uniquely from the kinetic energy spectra
of light charged particles.
Very interestingly, the hypothesis on prevalent break-up genesis of
$^3$He adopted to explain the high value of the average kinetic energy
is supported by the ability to reproduce its broad $dN/dK$ distribution and,
at the same time, validate the procedure used for the break-up reconstruction.
We mention here that the relatively low excitation of the source
and the numerous proofs in favor of existence of the break-up stage 
make earlier scenarios on $^3$He formation
from coalescing non-equilibrium nucleons 
or an expanding source less plausible.
The size of the average freeze-out volume and the corresponding
amount of collective flow
may be selected from the multiple solutions provided by
the break-up restoration algorithm performing
a consistency test with a
multifragmentation model with identical fragment definition.
Acceptable agreement between the MMM results and
predictions of the reconstruction algorithm
was obtained for $V=4V_0$ assuming for primary fragments
a constant limiting temperature $\tau$=6.8 MeV which leads to
an average excitation energy per mass unit $\epsilon/A$=2.5 MeV, 
in agreement with the experimental evaluations of Ref. \cite{hudan}.


\begin{thebibliography}{99}

\bibitem{gulminelli} F. Gulminelli, Ph. Chomaz, O. Juillet, M. J. Ison and
                     C. O. Dorso, in
		     {\em Proceedings for VI Latin American Symposium on
		       Nuclear Physics and Applications, Iguazu, Argentina (2005)}, 
		     to be published in Acta Phys. Hung. A (nucl-th/0511012).
\bibitem{frankland1} J. D. Frankland {\it et al.}, Nucl. Phys. {\bf A689}, 905 (2001).
\bibitem{frankland2} J. D. Frankland {\it et al.}, Nucl. Phys. {\bf A689}, 940 (2001).
\bibitem{hudan} S. Hudan {\it et al.}, Phys. Rev. C {\bf 67}, 064613 (2003).
\bibitem{piantelli} S. Piantelli {\it et al.}, Phys. Lett. {\bf B627}, 18 (2005).
\bibitem{charity} R. J. Charity, Phys. Rev. C {\bf 58}, 1073 (1998).
\bibitem{nicolas} N. Le Neindre, These de doctorat, Universit\'e de Caen (1999),
                  http://tel.ccsd.cnrs.fr/tel-00003741;
		  S. Hudan, These de doctorat, Universit\'e de Caen (2001).
\bibitem{bogatin} V. I. Bogatin {\it et al.}, Yad. Fiz. {\bf 32}, 1363 (1980).
\bibitem{gutbrod} H. H. Gutbrod {\it et al.}, Phys. Rev. Lett. {\bf 37}, 667 (1976). 
\bibitem{markiel} W. Markiel {\it et al.}, Nucl. Phys. {\bf A485}, 445 (1988).
\bibitem{poggi} G. Poggi {\it et al.}, Nucl. Phys. {\bf A586}, 755 (1995).
\bibitem{neubert} W. Neubert and A. S. Botvina, Eur. Phys. J. A {\bf 7}, 101 (2000).
\bibitem{bougault} R. Bougault {\it et al.},  in {\it Proceedings of the
XXVII International Workshop on Gross Properties of Nuclei and Nuclear Excitations,
Hirschegg, 1999}, edited by H. Feldmeier, J. Knoll, W. Noerenberg and J. Wambach
(GSI Darmstadt, 1999), p. 24.
\bibitem{ees} W. A. Friedman, Phys. Rev. C {\bf 42}, 667 (1990). 
\bibitem{tlim} Ad. R. Raduta, E. Bonnet, B. Borderie, N. Le Neindre and M. F. Rivet,
               Phys. Rev. C {\bf 72},057603 (2005).
\bibitem{rivet} M. F. Rivet {\it et al.}, Phys. Rev. C {\bf 25}, 2430 (1982).
\bibitem{vaz} L. Vaz {\it et al.}, Z. Phys. A {\bf 311}, 89 (1983).	       
\bibitem{li} Bao-An Li, D. H. E. Gross, V. Lips and H. Oeschler, Phys. Lett. {\bf B335},
	     1 (1994).
\bibitem{sangster} T. C. Sangster, M. Begemann-Blaich, Th. Blaich, H. C. Britt,
		   L. F. Hansen, M. N. Namboodiri and G. Peilert,
		   Phys. Rev. C {\bf 51} (1995).
\bibitem{bb} M. Begemann-Blaich {\it et al.}, Phys. Rev. C {\bf 58}, 1639 (1998). 
\bibitem{statmod} D. H. E. Gross, Rep. Progr. Phys. {\bf 53}, 605 (1990);
                  J. P. Bondorf, A. S. Botvina, A. S. Iljinov, I. N. Mishustin and K. Sneppen,
                  Phys. Rep. {\bf 257}, 133 (1995);
	          S. E. Koonin and J. Randrup, Nucl. Phys. A {\bf 474}, 173 (1987).
\bibitem{dynmod} M. Parlog, G. Tabacaru, J.P. Wieleczko, J.D. Frankland, B. Borderie,
                 A. Chbihi, M. Colonna, M.F. Rivet,
		 Eur. Phys. J. A {\bf 25}, 223 (2005) and references therein.
\bibitem{mmm} Al. H. Raduta and Ad. R. Raduta, Phys. Rev. C {\bf 55}, 1344 (1997); 
              {\it ibid.}, Phys. Rev. C {\bf 65}, 054610 (2002).
\bibitem{radii} Landolt-Bornstein, A. M. Hellwege , K. H. Hellwege (Eds.),
                Numerical data and functional relationships in science and technology, vol. I,
		Springer-Verlag, Berlin, 1961.
\bibitem{lefevre}  A. Le Fevre {\it et al.}, Nucl. Phys. {\bf A735}, 219 (2004).
\bibitem{de} J. N. De, S. K. Samaddar, X. Vinas and  M. Centelles,
	     Phys. Lett. {\bf B638}, 160 (2006). 	      
\bibitem{iljinov} A. S. Iljinov, {\it et al.}, Nucl. Phys. {\bf A543}, 517 
                  (1992).
\bibitem{natowitz}J. B. Natowitz, R. Wada, K. Hagel, T. Keutgen, M. Murray, A. Makeev,
                  L. Qin, P. Smith, and C. Hamilton, Phys. Rev. C {\bf 65}, 034618 (2002)
		  and references therein. 
\bibitem{baldo} M. Baldo, L. S. Ferreira and O. E. Nicotra,
                Phys. Rev. C {\bf 69}, 034321 (2004). 
\bibitem{npa2002} Al. H. Raduta and Ad. R. Raduta,
                     Nucl. Phys. {\bf A703}, 876 (2002).
\bibitem{mmm+smf} A. H. Raduta, M. Colonna, V. Baran, and M. Di Toro, 
                  Phys. Rev. C {\bf 74}, 034604 (2006).		  
\bibitem{smf} M. Colonna {\it et al.}, Nucl. Phys. {\bf A642}, 449 (1998). 
\bibitem{randrup-tlim} S. E. Koonin and J. Randrup, Nucl. Phys. {\bf A474}, 173 (1987).	
	  
\end{thebibliography}
\end{document}